# Continuous symmetry of $C_{60}$ fullerene and its derivatives

E.F.Sheka[1], B.S.Razbirin[2], E.A.Nikitina[3,4], D.K.Nelson[2]

[1] *Peoples' Friendship University of the Russian Federation, 117198 Moscow, Russia*
[2] *A.F. Ioffe Physical-Technical Institute, RAS, 194021 St.Petersburg, Russia*
[3] *Photochemistry Center RAS, 119421 Moscow, Russia*
[4] *Institute of Applied Mechanics RAS, 119991 Moscow, Russia*

**Abstract**. Conventionally, the $I_h$ symmetry of fullerene $C_{60}$ is accepted which is supported by numerous calculations. However, this conclusion results from the consideration of the molecule electron system, of its odd electrons in particular, in a close-shell approximation without taking the electron spin into account. Passing to the open-shell approximation has lead to both the energy and the symmetry lowering up to $C_i$. Seemingly contradicting to a high-symmetry pattern of experimental recording, particularly concerning the molecule electronic spectra, the finding is considered in the current paper from the continuous symmetry viewpoint. Exploiting both continuous symmetry measure and continuous symmetry content, was shown that formal $C_i$ symmetry of the molecule is by 99.99% $I_h$. A similar continuous symmetry analysis of the fullerene monoderivatives gives a reasonable explanation of a large variety of their optical spectra patterns within the framework of the same $C_1$ formal symmetry exhibiting a strong stability of the $C_{60}$ skeleton.

## 1. Introduction

Fullerene $C_{60}$ is related to those rare molecules, whose structure had been predicted before the molecule was discovered experimentally. The first suggestion was made by Jones in 1966 who assumed that intercalation of pentagon defects into a plain graphite layer (graphene according to the nowadays nomination) consisting of perfect hexagons would result in the transformation of this layer into a closed hollow shell that presents a giant carbon molecule [1]. In 1970 Osawa published a short communication [2] showing a possibility of existing $C_{60}$ molecule in the form of truncated icosahedron, consisting of 60 atoms. Next year Osawa and Yoshida discussed possible aromatic properties of the molecule [3]. Two years later Bochvar and Galpern [4] as well as Stankevich and coworkers [5] performed calculations of the molecule electronic structure. This calculation repeated in a few years by Davidson [6], showed that the closed-hollow-cell $C_{60}$ molecule is characterized by a large distance between the energies of the highest occupied (HOMO) and lowest unoccupied (LUMO) molecular orbitals, which points to a chemical stability.

The first discovering of the molecule is usually connected with the names of Kroto, Curl, and Smalley (KCS) [7] although a year earlier its presence among carbon clusters was heralded [8]. Not knowing the above cited results, KCS suggested a truncated icosahedron shape of the molecule, consisting of 32 faces, 60 apices (carbon atoms), and 90 ribs. Supposing a full equivalence of atom positions, the symmetry of the polyhedron as a geometrical replica for equilibrium positions of the molecule atoms was attributed to the point-group symmetry $I_h$. This suggestion has laid the foundation of a doubtless view on the molecule shape symmetry that has been supported by now by the absolute majority of the fullerene community in spite of a number of contradictions to this paradigm when interpreting some of the molecule properties. All the inconsistencies with the high symmetry are usually considered in terms of "a slight lowering of the molecule symmetry from the ideal symmetry $I_h$ under particular conditions". In the current paper we advocate a new consideration of the molecule property showing that there are serious reasons in the ground of these contradictions so that the symmetry problem of fullerene $C_{60}$ is not so simple.

## 2. $C_{60}$ shape and symmetry: Structural experiments

As accepted, the most reliable atomically mapped description of the molecule structure follows from the gas phase electron diffraction (EGD) [9]. Actually, the diffraction pattern is well fitted when suggesting that the molecule shape is a truncated icosahedron of the point-group symmetry $I_h$, which is formed by two types of C-C bonds, ones of a substantial double bond character and of $h$=1.398(10) Å in length while the other are of $p$=1.455(6) Å long and have a prevalent single bond character. Sixty carbon atoms are arranged in 20 six-membered and 12 five-membered rings. The C-C bonds separating two hexagons are double bonds ($h$) while the pentagon C-C bonds ($p$) are single. Under the same suggestion a good fitting to experimental data related to neutron diffraction (ND) from $C_{60}$ powder [10] and X-ray data (XRD) for the $C_{60}$ crystal [11] was obtained. The available data concerning C-C bond length are collected in Table 1. Analyzing the data, one can conclude that EGD provides the most accurate bond length determination whilst its accuracy is of $10^{-2}$Å only [9, 14]. At the same time, ND and XRD methods do not allow determining the value better than $10^{-1}$Å.

## 3. $C_{60}$ shape and symmetry: Quantum-chemical calculations

The modern Born-Oppenheimer-approached quantum chemistry can provide the accuracy in the bond length determination not worth than of $10^{-5}$Å. That is why quantum-chemical calculations can be regarded as highly accurate structure determination technique under conditions when a complete set of calculated results fits experimental findings of the object under consideration. Starting from [15, 16], the molecule has been repeatedly and thoroughly studied computationally (see [17-20] and references therein). In some sense, $C_{60}$ turned out to be a proving ground for testing different computational techniques, from a simplest to the most sophisticated. However, all calculations were based on the aromaticity concept which revealed itself in a close-shell or restricted approximation. All calculations have shown $I_h$-symmetry of truncated icosahedron structure to be corresponded to the lowest potential energy minimum in the singlet state. However, as was thoroughly discussed in [13, 21], the length of the molecule long C-C bonds exceeds the region that provides a complete covalent bonding of odd electrons, so that a remarkable difference $\Delta E^{RU}$ in the energies of restricted and unrestricted solutions should be expected highlighting non-stability of the restricted solutions and positioning a stable pure spin and physically real ground state of the molecule at lower energy. This is so indeed and Table 2 presents data based on single-determinant Hartree-Fock (HF) calculations. To show that the same happens for other fullerenic structure, data for $C_{70}$ and $Si_{60}$ molecules are added to the table.

As seen from the table, spin-contaminated energies of the unrestricted broken symmetry HF (UBS HF) solutions and pure spin energies of all three molecules are much lower than those of the restricted solutions. The $\Delta E^{RU}$ values constitute 1.53, 3.85, and 22.9% the RHF heats of formations for $C_{60}$, $C_{70}$, and $Si_{60}$, respectively. The finding points that the restricted solution is unstable in this case while unrestricted solution with lower energy is evidently closer to the physical reality. However, the energy lowering exhibits not only the instability of the restricted solution but is followed by the symmetry lowering as well. A tight connection between the energy lowering and the molecule symmetry has been thoroughly discussed in [25-32]. Since that it is commonly believed that the energy lowering is generally accompanied by a descent of symmetry (which adds geometry and/or symmetry instability to the spin instability). That is why lowering $I_h$ symmetry of RHF solution of $C_{60}$ to $C_i$ symmetry of the UBS HF one does not look quite unexpected.

Since the real state of the molecule is spin pure and since its energy is lower than that of spin-contaminated UBS HF one, the real symmetry of the molecule cannot be higher than the symmetry of the UBS HF state. That is why the UBS HF symmetry of the species is suggested for pure spin states in Table 2. A final decision on the real symmetry of the molecule might be put on shoulders of experiments. However, this way turned out to be rather complex and ambiguous. The situation is not particularly typical for fullerene species and is well known for other cases such as a discussion about $D_{6h}$ and/or $D_{3h}$ symmetry of the benzene molecule [33].

In the case of $C_{60}$, the statement that the point group symmetry of the molecule is $C_i$ seems to drastically contradict to the common opinion. Actually, a beautiful geometrical truncated-icosahedron shape seems do not allow even thinking about the molecule symmetry lower that $I_h$. And this intuitive expectation is considered to be proved by direct structural experiments and quantum chemical calculations. But any experimental proof has never been an absolute one providing a rigid "yes/no" answer concerning physical object symmetry due to unavoidable uncertainties inherited in experimental data. The latter is related even to such sophisticated technique as quantum chemical calculations where, as seen from Table 2, application of restricted approach instead of unrestricted one results in erroneous lifting of the molecule symmetry.

At the same time, the analysis of the molecule structure corresponding to either RHF or UBS HF solutions makes it possible to manifest what means the $C_{60}$ symmetry lowering from $I_h$ to $C_i$. Both approaches support the molecule truncated icosahedron shape formed by two groups of C-C bonds of different character. Thus, short bonds are close to the double bonds, which might be characterized by the Wiberg bond index [34]. In both cases the average Wiberg index for the bonds is 1.494 similar to that in benzene. As for long bonds, the average Wiberg index of 1.10 clearly evidences the single bond character. Therefore, the symmetry lowering is not connected with changing either the molecule shape or C-C bond character and is concerned with more delicate quantitative characteristics. Such is indeed the case, and the changing concerns the scattering of C-C bonds lengths. Figure 1 presents the dispersion of C-C bond distribution for two molecular structures that correspond to RHF and UBS HF solutions. As seen from the figure, if the average values of the bond lengths practically coincide for long bonds and slightly differ for short ones, the corresponding dispersions differ many times (four-time and 16-time for long and short bonds, respectively). This large difference in the bond length dispersion strongly actualizes the question concerning uncertainty of structural experiments performed since the two symmetry-different structures can be distinguished if only the accuracy of the bond length determination is better than $10^{-2}$Å. As for the most accurate EGD experiments [9, 14], their accuracy is close to the limit. However, a great number of adjustable parameters used under interpreting the data attenuate a confidence in the objective persuasiveness of the conclusion in favor of $I_h$ symmetry. A precise determination of the molecular geometry by neutron scattering from $C_{60}$ powder [10] gives values $h=1.391\pm0.063$Å and $p=1.452\pm0.066$Å that are even less accurate from the experimental viewpoint. The same can be addressed to the XRD data for the $C_{60}$ crystal [11]. Therefore, structural experiments are really not discriminative for the case and cannot resolve structures with either $I_h$ or $C_i$ symmetry.

Electronic optical spectra are another source of information related to the point-group symmetry. An exhausted and detailed analysis of highly structure-sensitive low-energy $S_0 \rightarrow S_1$ electronic optical spectra of $C_{60}$ forced the authors of review [35] to conclude that in spite of a high-symmetry pattern of the spectra as a whole, some peculiarities, such as a weak pure electronic $0_0^0$ transition in both absorption and fluorescence spectra, the vibronic series in the latter spectrum based on $g$-symmetry vibrations, absolutely allowed pattern of the phosphorescence spectrum, the "silent" modes in Raman and IR spectra [36], and others are inconsistent with the high symmetry of the molecule and cast some suspicion on the point.

The conclusions made on the basis of spectral studies as well as discriminative inability of structural experiments could seemingly be ponderable arguments in favor of the point-group symmetry $C_i$ although a high-symmetry pattern of the molecule spectra require an additional explanation. This fact as well as other physical reality shows that the symmetry problem of fullerene $C_{60}$ cannot be solved in terms of exact symmetry. A further evidence of the problem actuality was obtained when studying optical spectra of fullerene $C_{60}$ monoderivatives. Figure 2 presents three pairs of specular absorption/luminescence spectra belonging to molecules $C_{60}$ and two its monoderivatives dissolved in crystalline toluene. The spectra manifest the Shpol'skii effect, which concerns fine structure of vibronic spectra of soluted molecules in crystalline matrices under particular conditions [37]. The monoderivatives differ by added atomic groups that are shown in Fig.3. A high-symmetry pattern of the pristine fullerene spectra in Fig.2*c* is obvious: a very low intensity of the pure electronic $0_0^0$ bands in both spectra and the dominance of vibronic series based on non-totally *u*-symmetric vibrations that is Herzberg-Teller (HT) pattern. This pattern is typical for high symmetry molecules with forbidden or slightly allowed electron transitions (see spectra of benzene and naphthalene for example [41]). It is usually accepted as a convincing experimental evidence of high symmetry of the molecular object under study. If one might expect the pattern for $C_{60}$ basing on its overestimated symmetry, the symmetry of both derivatives is evidently $C_1$, which does not imply any forbidness of electronic transitions, so that their spectra should drastically differ from that of pristine fullerene. As turned out, the reality does not follow the expectation completely. Actually, as seen in the figure, the spectra of species **I** are absolutely different from those of pristine fullerene, while the spectra of species **II** are quite similar to the latter. The difference between spectra of $C_{60}$ and **I** concerns not the detailed vibrational structure of the spectra, which is expected, but the shape of the total spectra. Thus, a high-symmetry HT pattern of the $C_{60}$ is substituted by the Frank-Condon (FC) pattern that is characterized by intense $0_0^0$ bands and vibronic series based on *g*-symmetric vibrations only [42]. Oppositely to the case, spectra of species **II** preserve the HT pattern and a close similarity to the spectra of $C_{60}$. The only difference concerns the vibronic structure itself due to an obvious difference in vibrational modes of both molecules. Therefore, a low symmetry molecule **II** is characterized by a high-symmetry pattern of its electronic spectra.

The finding clearly shows up that the problem concerning the symmetry of the $C_{60}$ skeleton is not simple. The notations $C_i$ or $C_1$ themselves do not mean much since the matter is evidently not about these exact "yes/no" symmetries but about how much they both differ from, say, the $I_h$-one. The situation is not unique in the molecular world but the exclusive position of the fullerene forces to draw a particular attention to the problem. A concept based on the continuous symmetry approach seems to be the most suitable tool for the problem elucidation.

**4. Continuous symmetry concept**

A highly heuristic view that symmetry can be and, in many instances, should be treated as a continuous "gray" property, and not as a "black or white" one which exists or does not exist, was introduced by Zabrodski, Peleg, and Avnir (ZPA) in 1992 [43, 44] and thoroughly elaborated further by Prof. Avnir team as quantitative tools for the continuous symmetry measure (CSM) [43, 45], continuous chirality measure (CCM) [46], and continuous shape measure (CShM) [47, 48]. However, the topic has not been widely accepted so far and still is of interest of a rather narrow group of scientific community, mainly chemical theoreticians [43-53], while the majority of practicing chemists and physicists still prefer "black and white" approach. Speaking about continuous symmetry, the ZPA approach does not restrict oneself by the object symmetry and shape but concerns mass or electronic distributions as well. It means that *continuous symmetry* should be determined not only for the object shape but also for other massifs of data that provide the description of molecular properties on matrix-element level [52, 54]. However, usually, the shape symmetry is considered as basic one that strictly governs all other object properties.

As thoroughly analyzed in [55], the exact symmetry conceptual inflexibility which demands "yes or no" answer bends a rich reality of stereochemistry to the dictates of a strict codex, behind which are hidden (1) the paradigm that exact symmetry is superior and therefore must set the rules and (2) the practical aspect that fully symmetric situations are easier to formulate theoretically. The strict codex of perfect symmetry is particularly devastating in the case of small deviations from symmetry that force a jump in the symmetry description, the magnitude of which is totally out of proportion from that deviation. At the same time the continuous symmetry concept makes it possible to replace a conventional question: "What is the object symmetry under such or other conditions?" by another one: "How much the symmetry of the object deviate from an ideal and/or reference one under such or other conditions?" If the answer to the latter can be quantified, one can expect to be able to substitute usually met statements like: "The NH group perturbs the electron system of $C_{60}$ molecule to a small extent" (with respect to the intensity of the lowest electronic transition in Fig. 2) by some quantitative structure-property correlations. The continuous symmetry approach can satisfy these expectations to the most extent.

Mathematically, the approach is based on the distance-function formulation that makes it possible to determine a set of useful quantitative characteristics [43, 46, 55-58]. The first is the *continuous symmetry measure* CSM that is a distance function commonly employed in the symmetry analysis. If $M$ is a structure composed of $n$ vertices (atoms) in an original configuration, $Q_i$, and $G$ is any symmetry point group, the amount of $G$-symmetric shape in $M$ is defined as [55]

$$Sy(M,G) = \frac{1}{nD^2} \sum_{i=j}^{n} \|Q_i - P_i\|^2 . \qquad (1)$$

Here $P_i$ are the searched corresponding points of the nearest $G$-perfectly symmetric configuration, and $D$ is a size normalization factor (the *rms* of all distances from the center of mass to the vertices), making $Sy$ size-invariant. The distance $Q_i - P_i$ is squared to avoid sign limitation, as is a common practice in distance-function formulations. The bounds of $Sy$ are 0 ($Q_i$ coincides with $P_i$; $M$ is perfectly symmetric) and 1 (the nearest symmetric structure coalesces onto a center point, the distance to which is 1). To make the values more sensitive to the geometry deviation, the latter are usually scaled by 100. Several algorithms were developed for the identification of $P_i$ and implemented in working programs SYMMETRY and SHAPE [59]. $Sy(M,G)$ is a quantitative description of CSM within 0-100 scale.

The *continuous symmetry numbers* (CSNs) is the next distance function which measures the deviation from rotational symmetry. Originally they were introduced as

$$\sigma = \sum_{k} r_k . \qquad (2)$$

to correct evaluation of rotational entropy [56]. Here $k$ numbers proper $C_n$ rotations and

$$r_k = 1 - Sy_k \qquad (3)$$

where $Sy_k$ evaluates the degree of a specific rotation operation $k$ from perfectness and is expressed as

$$Sy_k(M,k) = \frac{1}{nD^2} \sum_{i=j}^{n} \|Q_i - P_{ik}\|^2 . \qquad (4)$$

The distance function $Sy_k$ should be determined for each of $k$ symmetry elements of the point group $G$ within the interval from 0 to 1. Obviously, the last three equations can be expanded over improper rotations $S_n$, inversion ($S_2$), mirror symmetry $C_s$, and chirality $C_h$ thus allowing determination of *continuous symmetry level* (CSL) $\sigma_{cont}$ (the term was suggested by D.Avnir and C.Dryzhun and accepted by the authors with gratitude). When the set of symmetry elements of the point group $G$ is formed of the above elements, CSL $\sigma_{cont}$ describes an absolute contribution of the $G$-symmetry in the studied structure while $\eta_{cont} = \sigma_{cont}/\sigma_{classic}$, where $\sigma_{classic}$ equal to $k$ is a perfect measure of the $G$-symmetry, presents a relative contribution.

A complex of programs created in the Hebrew University of Jerusalem [59] offers a large possibility to determine a consistence of shapes of two structures by SHAPE program, to find a selected point group symmetry contribution into a studied structure by SYMMETRY program and to evaluate chirality of the structure by CHIRALITY program. Let us look at pristine fullerene $C_{60}$ and its derivatives from the viewpoint of continuous symmetry.

## 5. Fullerene $C_{60}$ and its monoderivatives

The first application of the continuous symmetry concept to fullerene $C_{60}$ concerned the symmetry of its anions from $C_{60}^{-1}$ to $C_{60}^{-6}$ [55]. Classically, the structural change induced in $C_{60}$ upon charging to $C_{60}^{-1}$ is associated with an abrupt drop in symmetry from $I_h$ to $D_{3h}$, which leads to a grossly overestimated rotational entropy change and to CSN $\sigma_{cont}$, limited by proper rotations, equal to 60 and 6, respectively. Similar abrupt changes without any reasonable order, as well as grossly overestimated values of rotational entropy changes, are also obtained for the rest of other anions toward $C_{60}^{-6}$. At the same time, changes in CSNs, calculated for rotations involved in the rotational subgroup of the point group $I_h$, for all anions ($\sigma_{cont}$ lies in the interval from 58.73 to 58.72) constitute only 2.2% instead of 90% expected classically. This means that the anion $C_{60}$ skeleton preserves the $I_h$-ness to a great extent.

In the current paper we proceed with the consideration of the $C_{60}$ fullerene symmetry and address the symmetry measure to answer two questions: 1) how much $C_i$- structure of the molecule, which follows from the UBS HF solution, deviates from the $I_h$-structure of the RHF solution and 2) how much $I_h$ symmetry contributes to structure of various $C_{60}$ monoderivatives as a whole and to their $C_{60}$ skeletons, in particular. Answering the first question, we use two continuous symmetry measures provided by programs SHAPE and SYMMETRY. In the first case, the corresponding symmetry measure $Sy(M,G)$ was obtained in due course of the comparison of the molecule $C_{60}$_RHF ($I_h$) and $C_{60}$_UBS HF ($C_i$) shapes. The CSM analysis was provided by the same fixed numeration of the molecule atoms in both structures. In its turn, $Sy_k$ values, provided by the SYMMETRY program, allowed evaluating the response of $C_i$- structure on each symmetry element of $I_h$ point group. The consideration of the fullerene derivatives was carried out in terms $Sy_k$.

The application of the shape symmetry analysis to $C_{60}$_RHF and $C_{60}$_UBS HF structures gives the $Sy(M,G)$ values equal to 0 and 0. 011951, when either identical or different structures are compared. If $Sy(M,G) = 0$ points to a trivial identity of the structure compared with oneself,

$Sy(M,G) = 0.011951$ (0-100 scale) evidences a practically negligible deviation of the $C_i$-structure from the $I_h$-one.

Eight symmetry elements of $I_h$ point group are given in the top of Table 3. All of them, beside symmetry planes σ, can be directly tackled within the SYMMETRY program to obtain corresponding $Sy_k$ values. As for the planes, those are perpendicular to $C_2$ axes, so that we can consider them via chirality's element $C_h$. The table contains symmetry measures $Sy_k$ determined according Eq. (4). The numbers are given in 0-1 scale.

The values related to $C_{60}$_RHF ($I_h$) are equal zero, as should be expected. $Sy_k$ values for $C_{60}$_UBS HF ($C_i$) differ from zero starting from the fourth and/or fifth digits after the point. However, the deviation is actually small that is why a summary $\sigma_{cont}$ constitutes 119.99 instead of $\sigma_{classic}$= 120 and $\eta_{cont} = 99.99\%$. Therefore, the $C_i$-structure is practically of $I_h$ symmetry. This finding correlates perfectly with the data of the shape analysis. A similar conclusion seems to be expected for the real symmetry of the $C_{60}$ molecule in the pure spin singlet state, which formally cannot be higher than the exact $C_i$ symmetry of the UBS HF solution but in terms of continuous symmetry should be predominantly $I_h$. The high continuous symmetry of the molecule provides high-symmetry patterns of all the symmetry sensitive experimental recordings. At the same time, its deviation from the exact $I_h$ symmetry may be used to explain all the inconsistencies of experimental recordings from the exact high-symmetry ones.

As for fullerene derivatives, the continuous symmetry measure analysis makes it possible to trace changes in both the derivative as a whole and its $C_{60}$ skeleton when an atomic group is added to the pristine fullerene. Let us look at the relevant data for the structures shown in Fig.3 which are presented in Table 3 and Table 4. The data related to the molecules themselves are bolded. As seen from the table, all additions considerably disturb the molecule structures so that their deviations from the $I_h$-one are quite large. Actually, $\sigma_{cont}$ in Table 4 changes from 113.128 to 86.129 resulting in $\eta_{cont}$ laying in the interval from 94.3 to 71.8%. In its turn, the expansion of the changes over symmetry elements in term of $Sy_k$ in Table 3 forms the ground for a new 'symmetry language' for the change description as well as for its comparative analysis with respect to varied structure and composition of the added groups. This makes it possible to substitute usually met a typical sentence such as "$CH_3$ unit causes a rather weak effect on the benzene ring structure" by another one: "$C_s$ symmetry of toluene molecule preserves 77% symmetry with respect to $C_6$ operation, 80% of $C_3$, 98% of $C_2$, 92.3% of inversion, 100% of $C_s$, thus conserving ~92% of $D_{6h}$ symmetry in total". Similarly, the data presented in Table 3 may be used to describe $C_1$ symmetry of the discussed derivatives. It is clearly from the table, that a concrete content of this description is different for the molecules characterized by different addends, in spite of the fact that the latter are added to the same atoms of the pristine fullerene cage.

Oppositely to the whole molecules, their $C_{60}$ skeletons having $C_1$ exact symmetry in all cases preserve (99.99÷99.98) % of $I_h$ symmetry, as follow from Tables 3 and 4. Therefore, the same molecule may produce both low-symmetry and high-symmetry experimental patterns depending on which atoms are involved in the empirical response. If those are skeleton atoms only, one obtains a high-symmetry patterned response. If addend atoms are mostly involved, a low-symmetry patterned response will be obtained. Let us consider this effect exemplified by optical electronic spectra of the studied derivatives.

## 6. Optical electronic spectra of fullerene $C_{60}$ and its derivatives

The considered electronic spectra correspond to optical transitions between the ground and excited electronic states produced within the first HOMO-LUMO gap [35]. In the framework of single-determinant HF approximation, the atomic function composition of HOMO and LUMO will govern the atomic-sensitive characteristic of the wave functions of excited states. Table 5 presents a percentage contribution of addend atoms in HOMOs and LUMOs of the studied derivatives. As follows from the table, the addend atoms do not contribute to LUMOs in all cases. As for HOMO contributions, the latter is zero for molecule **II**, constitutes 0.9÷1.4% for molecules **III**, **IV**, **I** and absolutely dominates at the level of 97.1% for species **V**. The data obtained allows for making certain conclusions concerning optical spectra of the molecules.

1. Spectra of molecule **II** are governed by the participation of $C_{60}$ skeleton atoms only so that the spectra pattern should be similar to that of the pristine fullerene.
2. Spectra of molecules **I**, **III**, and **IV** are initiated by both $C_{60}$ skeleton atoms and addends so that the whole molecule is responsible for the spectra intensity. Suggesting that the intensity would correlate with the deviation of the molecule structure from $I_h$-symmetry (taken as a measure of the $C_1$-ness in terms of $I_h$ symmetry), the intensity should increase when going from molecule **IV** (16%) to **III** (17%) and then to **I** (25%). At the same time intensity of spectra, which correspond to the excitation of the lowest excited state, should be compared with that of pristine fullerene due to low contribution of the addend atoms into the relevant HOMOs.
3. Spectra of molecule **V** should have a spectral-allowed pattern and be of the highest intensity due to both a high measure of the $C_1$-ness (19.9%) and a high addend atom contribution to the HOMO.

These predictions are supported by direct calculations of the oscillator strength related to optical dipole transitions to the first low lying excited states. The data obtained in the framework of ZINDO/S method [60] implemented in the GAUSSIAN package [61] for molecular structures shown in Fig. 3 are given in Table 6. The table lists five lowest excited states that contribute to low-frequency optical absorption spectra of the species [35]. As seen from the table, both $I_h$- and $C_i$ -structures of $C_{60}$ are characterized by zero oscillator strength of electronic transitions in this region. This might be interpreted that in spite of a formal allowance of the transitions for the $C_i$ molecule, its wave functions are in fact enough highly symmetrical to provide zero dipole matrix elements related to the transitions. This appears to be consistent with the symmetry analysis discussed earlier.

The remaining five molecules are characterized by allowed transitions in the region, summary intensity of which forms a series **V** > **I** ≥ **IV** > **III** > **II**. It should be noted that when summary intensities of the transitions of molecules **V**, **I**, and **IV** are comparable, those of species **III** and **II** are two-three times less. The series is fully synchronous with that one presenting the deviation of the molecule symmetry from the $I_h$ one as follows from Table 4. Actually, following the disclosed tendency **V** > **I** ≥ **IV** > **III**, one can see a synchronous directed changing in the spectra pattern of molecules **V**, **I**, **IV**, and **III** presented in Fig. 4. The spectra of first three molecules have a clearly exhibited FC vibronic pattern and their intensity somewhat decreases downward, accompanying with a simultaneous extension of the spectra vibronic structure of the luminescence spectra. Thus, in the case of molecule **V**, zero-vibration (pure electronic) $0_0^0$ band dominates in the spectrum while its vibrational repetitions are rather scarce. The same retains in the case of molecules **I** and **IV** with the domination of the $0_0^0$ band slightly decreased. Oppositely to the cases, spectra of molecule **III**, whose intensity differ twice from those of the above mentioned molecules, are of different shape; the $0_0^0$ band in both absorption and luminescence spectra constitutes an ordinary part of extended vibronic series that are a mixture of FC and HT series in the favor to the former so far starting with a rather weak but still clearly vivid $0_0^0$ band [31]. This dependence of spectra vibronic structure on the intensity of electronic transition is a typical feature in molecular spectroscopy of large molecules where weak electronic transitions are presented by extended series of vibronic bands oppositely to strong allowed transitions. This is connected with the fact that in the former case there is a comparable chance for a number of different vibrations to be exhibited, including both *g*- and *u*- symmetric vibrations. Oppositely, in the latter case, the vibrational repetitions are provided by a limited number of *g*-symmetric vibrations only, series extension of which is determined by the shift of the equilibrium positions of atoms under the electronic excitation. For large molecules the latter is usually small, which results in short vibronic series. Spectra of molecule **V** in Fig.4*a* clearly demonstrate the latter tendency, particularly, for the luminescence partner. Lowering intensity for molecules **I** and **IV** produces a vivid extension of the vibronic structure of their spectra (Figs.4*b* and 4*c*) while a further lowering the intensity of the electronic transitions for molecule **III** causes a substantial enrichment of the vibronic structure transforming it into a mixture of FC and HT series, so that it becomes quite similar to a rich vibronic HT structure of electronic transitions of the pristine fullerene $C_{60}$ spectrum.

Figure 5 presents spectra of molecules **III** and $C_{60}$ in more details based on fine-structured Shpolskii's spectra at 2K. Unfortunately, due to very strong tendency to clusterization (see for example [62]), molecules **II** cannot be distributed in the toluene matrix as individual entities, which prevents from obtaining their fine-structured vibronic spectra similar to those of molecules **III** and $C_{60}$ so that even at low temperature its spectra consists of rather broad bands (Fig. 5*c*). As clearly seen in Figure 5*a* and 5*b*, spectra of both molecules present a mixture of FC and HT series with the difference that when the FC series is the most intense in the former case, the HT series dominates in the latter. Lowering the absolute intensity when passing from molecule **III** to $C_{60}$ inhibits the intensity of the FC series. The same happens when going from molecule **III** to **II** since the spectra of the latter are just overlapping of those of $C_{60}$.

Basing on the finding it might be said that summary strength of the electronic transitions at the level between 0.005 and 0.006 which is characteristic for **II** and **III** (Table 6) can be considered as a limit for a transition from a predominantly forbidden HT spectra pattern to predominantly allowed FC one. In the studied case this corresponds to 10-15% deviation from the highest $I_h$ symmetry. Therefore, '(0.006-0.005)/(10-15)%' relationship seems to be a bordering condition when changing in structure-symmetry of a fullerene molecule will cause a qualitative reconstruction of its optical spectra that allows establishing a lifting and /or lowering of the molecule symmetry. Similar quantitative relationships can be established for other structure-sensitive experimental techniques such as NMR, IR and Raman spectroscopy, etc. Evidently, quantitative expressions for bordering cases might be different for different techniques.

## 7. Conclusion

Continuous symmetry approach may be suggested as a new 'symmetry language' that provides a quantitative description of a molecule structure in terms of higher symmetries and forms the grounds for establishing quantitative structure-property relationships. Optical spectra offer a good platform for the latter due to their high sensitivity to structure deviation. Applied to $C_{60}$ fullerene and its monoderivatives, the approach has proved its great efficiency and has highlighted a big stability of the fullerene skeleton, the deviation of whose symmetry from $I_h$ is rather small even with respect to non-symmetrical additions of various molecular units.

## Acknowledgement

One of the authors (E.Sh.) is grateful to S.S.Trach for attracting her attention to a quantitative approach for molecule symmetry consideration. A particular gratitude to D.Avnir, C.Dryzhin, and M.Pinsky for numerous fruitful discussions, for presenting a possibility to be acquainted with programs SHAPE and SYMMETRY, for giving permission as well as invaluable practical lessons of the programs use. The work was partially financially supported by the RFBR (grant № 07-03-00755).

References

1. Jones, D.E.H. Ariadne. *New Scientist.* 32: 245 (1966).
2. Osawa E. Kagaku (Kyoto) 25, 854 (1970).
3. Yoshida, Z.; Osawa, E. *Aromaticity*. (In Japanese).174. Kyoto: Kagaku Dojin. 1971.
4. Bochvar, D.A.; Galpern,E.G. Doklady Physics 209, 610 (1973).
5. Stankevich, I.V.; Nikerov, M.V.; Bochvar, D.A. Russ. Chem. Rev. 53, 640 (1984).
6. Davidson R.A. Theor. Chim. Acta. 58, 193 (1981).
7. Kroto H.; Heath J.R.; O'Brien S.C.; Curl R.F.; Smalley R.E. Nature 318, 354 (1985).
8. Rohlfing E.A.; Cox D.M.; Kaldor A. J.Chem. Phys. 81, 3322 (1984).
9. Hedberg, K.; Hedberg, L.;Bethune, D.S.; Brown, C.A.; de Vries, M.; Dorn, H.C.; Johnson, R.D. Science 254, 410 (1991).
10. Leclercq,F.; Damay, P; Foukani, M; Chieux, P.; Bellisent-Funnel, M.C.; Rassat, A.; Fabre, C. Phys. Rev. 48B, 2748 (1993).
11. Slovokhotov, Yu.L.; Moskaleva, I.V.; Shil'nikov, V.I.; Valeev, E.F.; Novikov, Yu.N.; Yanovski, A.I. ; Struchkov, Yu.T. Mol.Mat. 8, 117 (1996).
12. Yanonni C.S., Bernier, P.P. ; Bethune, D.S. ; Meijer, G. ; Salem, J.K. J.Am.Chem.Soc. 113, 3190 (1991).
13. Sheka, E.F. Int.Journ.Quant.Chem. 107, 2361 (2007).
14. Hedberg, L.; Hedberg, K.; Boltalina, O.V. ; Galeva, N.A.; Zapolskii, A.S.; Bagryantsev, V.F. J.Phys.Chem. 108 A, 4731 (2004).
15. Chang, A.H.H.; Ermler, W.C.; Pitzer R.M. J. Phys. Chem. 95, 9288 (1991).
16. Weaver, J.H.: Acc. Chem. Res. 25, 143 (1992).
17. Lee, B.X.; Cao, P.L. ; Que, D.L. Phys Rev 61B, 1685 (2000).
18. Nagase, S. Pure Appl. Chem. 65**,** 675 (1993).
19. Slanina,  Z., Lee, S.L.: Fullerene Sci.Technol.2 459 (1994).
20. Lee, B.X.; Jiang, M.; Cao, P.L. J.Phys. Condens. Matter 11**,** 8517 (1999).
21. Sheka, E.F., Chernozatonslii, L.A. J. Phys. Chem. C, 111, 10771 (2007).
22. Zayets, V.A. "CLUSTER-Z1: Quantum-Chemical Software for Calculations in the s,p-Basis",  Institute of Surface Chemistry, Nat. Ac.Sci. of Ukraine: Kiev, 1990.
23. Weaver, J.H.; Martins, J.L.; Komeda, ; Chen, Y.; Ohno, T.R.; Kroll, G.H.;
24. Wang, X.-B.; Ding, C.-F.; Wang, L.-S. J.Chem.Phys. 110, 8217 (1999).
25. Overhauser, A.W.  Phys.Rev.Letts. 4, 415, 466 (1960).
26. Touless, D.J. The Quantum Mechanics of Many-Body Systems. AP: NY, 1961
27. Adams, W.H. Phys.Rev. 127, 1650 (1962).
28. Löwdin, P.O. Phys. Rev. 97, 1509 (1955).
29. Koutecki, J. J.Chem.Phys. 46, 2443 (1967).
30. Čižek, J.;  Paldus, J. J.Chem.Phys. 47, 3976 (1967).
31. Benard, M. J.Chem.Phys. 71, 2546 (1979).
32. Paldus, J.; Veillard, A. Mol. Phys. 35, 445 (1978).
33. Ermer, O. Angew. Chem. Int. Ed. Engl. 26, 782 (1987).
34. Wiberg, K.B. Tetrahedron 24, 1083 (1968).
35. Orlandi, G.; Negri, F. Photochem. Photobiol. Sci. 1, 289 (2002).
36. Horoyski, P.J.; Thewalt, M.L.W.; Anthony, T.R. Phys. Rev. B 54, 920 (1996).
37. Shpol'skii, E.V. Soviet Physics-Uspekhi 77, 321 (1962).
38. Razbirin B.S.; Starukhin A.N.; Nelson D.K.; Sheka E.F.; Prato M. Int. Journ. Quant. Chem. 107, 2787 (2007).
39. Razbirin B.S.; Sheka E.F.; Starukhin A.N.; Nelson D.K.; Troshin P.A.; Lyubovskaya R.N. Phys. Sol. State 51, 1315 (2009).
40. Sheka E. F.; Razbirin B. S.; Starukhin A. N.; Nelson D. K., Degunov M. Yu.; Fazleeva G. M.; Gubskaya V. P.; Nuretdinov I. A. Phys. Sol. State 51, 2193 (2009).

41. Broude, V.L.; Rashba, E.I.; Sheka, E.F. Spectroscopy of Molecular Excitons. Springer: Berlin (1985).
42. Osadko, I.S. Physics-Uspekhi, 22, 311 (1979).
43. Zabrodski H.; Peleg, S.; Avnir, D. J. Amer. Chem. Soc. 114, 7843 (1992).
44. Zabrodski H.; Peleg, S.; Avnir, D. J. Amer. Chem. Soc. 115, 8278 (1993)
45. Pinsky, M.; Dryzun, C.; Casanova, D.; Alemany P.; Avnir, D. J. Comp. Chem. 29, 2712 (2008).
46. Zabrodsky, H.; Avnir, D. J. Am. Chem. Soc. 117, 462 (1995).
47. Pinsky, M.; Avnir, D. Inorg. Chem. 37, 5575 (1998).
48. Alvarez, S.; Avnir, D.; Llunell, M.; Pinsky, M. New J. Chem. 26, 996 (2002).
49. Alikhanidi, S.; Kuz'min, V. Zh. Strukt. Chimii 1998, 39, 548.
50. Alikhanidi, S.; Kuz'min, V. J.Mol.Mod. 1999, 5, 116/
51. Tratch, S. S.; Zefirov, N. S. J. Chem. Inf. Comp. Sci. 38, 331 (1998).
52. Grimm, S. Chem. Phys. Lett. 1998, 297, 15.
53. Petitjean, M. Entropy 5, 271 (2003).
54. Avnir, D.; Dryzhun, C. Phys. Chem. Chem. Phys. 11, 9653 (2009).
55. Estrada, E.; Avnir, D. J. Amer. Chem. Soc. 125, 4368 (2003).
56. Avnir, D. ; Katzenelson, O. ; Keinan, S. ; Pinsky, M. Salomon, Y.; Zabrodsy. In Concept in Chemistry. A Contemporary Challenge; Rouray, D. Ed. John Wiley aqnd Sons: New York, p.283 (1997).
57. Keinan, S.; Avnir, D. J. Amer. Chem. Soc. 120, 6152 (1998).
58. Katzenelson, O.; Avnir, D. Chem.-Eur. J. 6, 1346 (2000).
59. SHAPE, SYMMETRY, and CHIRALITY programs. (2010). http://www.csm.huji.ac.il/new/
60. M. Zerner, Reviews in Computational Chemistry, Volume 2, Eds. K. B. Lipkowitz and D. B. Boyd, VCH, New York, 313, (1991).
61. Frisch, M. J.; Trucks, G. W.; Schlegel, H. B.; Scuseria, G. E.; Robb, M. A.; Cheeseman, J. R.; Montgomery, Jr., J. A.; Vreven, T.; Kudin, K. N.; Burant, J. C.; Millam, J. M.; Iyengar, S. S.; Tomasi, J.; Barone, V.; Mennucci, B.; Cossi, M.; Scalmani, G.; Rega, N.; Petersson, G. A.; Nakatsuji, H.; Hada, M.; Ehara, M.; Toyota, K.; Fukuda, R.; Hasegawa, J.; Ishida, M.; Nakajima, T.; Honda, Y.; Kitao, O.; Nakai, H.; Klene, M.; Li, X.; Knox, J. E.; Hratchian, H. P.; Cross, J. B.; Bakken, V.; Adamo, C.; Jaramillo, J.; Gomperts, R.; Stratmann, R. E.; Yazyev, O.; Austin, A. J.; Cammi, R.; Pomelli, C.; Ochterski, J. W.; Ayala, P. Y.; Morokuma, K.; Voth, G. A.; Salvador, P.; Dannenberg, J. J.; Zakrzewski, V. G.; Dapprich, S.; Daniels, A. D.; Strain, M. C.; Farkas, O.; Malick, D. K.; Rabuck, A. D.; Raghavachari, K.; Foresman, J. B.; Ortiz, J. V.; Cui, Q.; Baboul, A. G.; Clifford, S.; Cioslowski, J.; Stefanov, B. B.; Liu, G.; Liashenko, A.; Piskorz, P.; Komaromi, I.; Martin, R. L.; Fox, D. J.; Keith, T.; Al-Laham, M. A.; Peng, C. Y.; Nanayakkara, A.; Challacombe, M.; Gill, P. M. W.; Johnson, B.; Chen, W.; Wong, M. W.; Gonzalez, C.; and Pople, J. A. *Gaussian 03* (*Revision B.03*); Gaussian Inc., Pittsburgh PA (2003).
62. Sheka, E.F.; Razbirin, B.S.; Starukhin, A.N.; Nelson, D.K.; Degunov, M.Yu.; Troshin, P.A.; Lyubovskaya, R.N. J. Nanophot. SPIE 3, 033501 (2009)

**Table 1** Experimental data on C-C bond lengths of the $C_{60}$ molecule, Å

| Experiment[1)] | Single C-C bonds ($p$) | Double C-C bonds ($h$) | Reference |
|---|---|---|---|
| EGD ($I_h$) | 1.455(6) | 1.398(10) | [9] |
| ND (powder) ($I_h$) | 1.452(66) | 1.391(63) | [10] |
| XRD ($I_h$) | 1.452(66) | 1.391(63) | [11] |
| $^{13}$C NMR ($I_h$) | 1.450 (15) | 1.400 (15) | [12] |
| QCh (UBS HF AM1) ($C_i$) | 1.464(13) | 1.391(32) | [13] |
| QCh (RHF AM1) ($I_h$) | 1.463(3) | 1.385(0.4) | [13] |

Note 1: Uncertainties of physical experiments and dispersions of the C-C bond distributions provided by UBS HF AM1 calculations are given in parentheses.

**Table 2** Characteristics of singlet ground state of fullerenes [13] [1].

| | Solution / Data | $C_{60}$ | $C_{70}$ | $Si_{60}$ |
|---|---|---|---|---|
| RHF | Heat of formation [2], *kcal/mol* | 970.180 | 1061.136 | 1295.967 |
| | Symmetry | $I_h$ | $D_{5h}$ | $C_i$ |
| | Number of effectively unpaired electrons, *e* | 0 | 0 | 0 |
| | Total spin $\langle \hat{S}^2 \rangle$ | 0 | 0 | 0 |
| | Ionization potential [3], *I*, eV | 9.64 | 9.14 | 8.00 |
| | Electron affinity [3], $\varepsilon$, eV | 2.95 | 3.27 | 3.38 |
| UBS HF | Heat of formation [2], *kcal/mol* | 955.362 | 1020.226 | 1011.722 |
| | Symmetry | $C_i$ | $D_{5h}$ | $C_1$ |
| | Number of effectively unpaired electrons, *e* | 9.84 | 14.40 | 63.52 |
| | Total spin $\langle \hat{S}^2 \rangle$ | 4.92 | 7.20 | 31.76 |
| | $\Delta E^{RU}$, *kcal/mol* | 14.818 | 40.910 | 284.245 |
| | Ionization potential [3], *I*, eV | 9.87 (8.74[a]) | 9.87 | 8.98 |
| | Electron affinity [3], $\varepsilon$, eV | 2.66 (2.69[b]) | 2.73 | 2.70 |
| Pure spin state | Heat of formation, *kcal/mol* | 899.580 | 963.176 | 994.597 |
| | Symmetry | $C_i$ | $D_{5h}$ | $C_1$ |
| | Number of effectively unpaired electrons, *e* | 0 | 0 | 0 |
| | Total spin $\langle \hat{S}^2 \rangle$ | 0 | 0 | 0 |
| | $\Delta E^{RPS}$ [4], *kcal/mol* | 70.600 | 97.960 | 314.903 |

[1] Data were obtained by using RHF and UHF versions of the AM1 semi-empirical technique implemented in the CLUSTER-Z1 software [22]. UHF presents a single-determinant broken symmetry unrestricted Hartree-Fock approach.

[2] Molecular energies are presented by heats of formation $\Delta H$ determined as $\Delta H = E_{tot} - \sum_A (E_{elec}^A + EHEAT^A)$. Here $E_{tot} = E_{elec} + E_{nuc}$, while $E_{elec}$ and $E_{nuc}$ are the electron and core energies. $E_{elec}^A$ and $EHEAT^A$ are electron energy and heat of formation of an isolated atom, respectively.

[3] Here ionization potential and electron affinity correspond to energies of HOMO and LUMO, respectively, just inverted by sign. Experimental data for relevant orbitals of $C_{60}$ are taken from [23] (a) and [24] (b).

[4] $\Delta E^{RPS}$ presents the difference between the heats of formation of RHF and pure spin states.

**Table 3** Continuous symmetry measures $Sy$ related to element symmetry of $I_h$ point group [1)]

| Molecule | E | C5 | C3 | C2 | i | S10 | S6 | σ (Ch) |
|---|---|---|---|---|---|---|---|---|
| C60_RHF | 0 | **0** | **0** | **0** | **0** | **0** | **0** | **0** |
| C60_UBS HF | 0 | **0.000093** | **0.000085** | **0.000054** | **0.000056** | **0.00011** | **0.0001** | **0.000056** |
| I | 0 | **0.097742** | **0.058939** | **0.374478** | **0.945437** | **0.47872** | **0.42051** | **0.033549** |
| I_sk | 0 | 0.000231 | 0.000186 | 0.000128 | 0.000132 | 0.00024 | 0.00023 | 0.000132 |
| II | 0 | **0.040622** | **0.034247** | **0.0007** | **0.064343** | **0.14461** | **0.06438** | **0.025276** |
| II_sk | 0 | 0.000094 | 0.000086 | 0.000057 | 0.000058 | 0.0001 | 0.00011 | 0.000056 |
| III | 0 | **0.041919** | **0.050337** | **0.036587** | **0.363893** | **0.39697** | **0.39599** | **0.0193** |
| III_sk | 0 | 0.000102 | 0.000092 | 0.000067 | 0.00006 | 0.00011 | 0.00011 | 0.00006 |
| IV | 0 | **0.056269** | **0.032465** | **0.032465** | **0.987355** | **0.35785** | **0.35683** | **0** |
| IV_sk | 0 | 0.000215 | 0.000159 | 0.000121 | 0.000121 | 0.00023 | 0.00022 | 0.000121 |
| V | 0 | **0.101112** | **0.068744** | **0.042874** | **0.877885** | **0.63536** | **0.5452** | **0.159748** |
| V_sk | 0 | 0.000225 | 0.000165 | 0.000126 | 0.000127 | 0.00024 | 0.00023 | 0.000127 |

[1)] Equilibrium structures of species molecules are given in Fig.3. Addition 'sk' marks 60-atom fullerene skeleton of the corresponding derivatives.

**Table 4** Classical and continuous symmetry numbers of fullerene $C_{60}$ and its monoderivatives [1)]

| Molecule | Symmetry | σ classic | σ continuous | | |
|---|---|---|---|---|---|
| | | | E + Cn | i + Sn + σ | Total |
| C60_RHF | *Ih* | 120 | **60** | **60** | **120** |
| C60_UBS HF | *Ci* | 2 | **59.995** | 59.994 | 119.99 |
| I | *C1* | 1 | 50.858 | 38.652 | 89.51 |
| I_sk | *C1* | 1 | **59.989** | 59.987 | **119.976** |
| II | *C1* | 1 | 58.33 | 54.798 | 113.1279 |
| II_sk | *C1* | 1 | **59.995** | 59.994 | **119.989** |
| III | *C1* | 1 | 57.438 | 41.899 | 99.338 |
| III_sk | *C1* | 1 | **59.995** | 59.994 | **119.989** |
| IV | *C1* | 1 | 57.513 | 43.288 | 100.801 |
| IV_sk | *C1* | 1 | **59.99** | 59.988 | **119.978** |
| V | *C1* | 1 | 55.555 | 30.573 | 86.129 |
| V_sk | *C1* | 1 | **59.989** | 59.988 | **119.977** |

[1)] See footnote to Table 3.

**Table 5.** Percentage contribution of the addend atoms into HOMOs and LUMOs[1]

| Molecule | HOMO | LUMO |
|---|---|---|
| **I** | 1.4 | 0 |
| **II** | 0 | 0 |
| **III** | 0.9 | 0 |
| **IV** | 1.0 | 0 |
| **V** | 97.1 | 0 |

[1] See footnote 1 to Table 3

**Table 6.** Calculated energies of the lowest excited states and oscillator strengths of the lowest electronic transitions (ZINDO/S)

| Molecule | Energy, eV | Oscillator strength |
|---|---|---|
| $C_{60}$-RHF | 2.2394 | 0 |
|  | 2.2414 | 0 |
|  | 2.2429 | 0 |
|  | 2.3156 | 0 |
|  | 2.3163 | 0 |
|  | Σ | **0** |
| $C_{60}$-UBS HF | 2.0785 | 0 |
|  | 2.1456 | 0 |
|  | 2.1458 | 0 |
|  | 2.2217 | 0 |
|  | 2.2856 | 0 |
|  | Σ | **0** |
| I | 1.9712 | 0.0098 |
|  | 1.9898 | 0.0004 |
|  | 2.1897 | 0.0001 |
|  | 2.1897 | 0.0001 |
|  | 2.2506 | 0.0003 |
|  | Σ | **0.0107** |
| II | 2.0120 | 0 |
|  | 2.0490 | 0.0043 |
|  | 2.1440 | 0 |
|  | 2.225 | 0.0003 |
|  | 2.3 | 0.0001 |
|  | Σ | **0.0047** |
| III | 2.0548 | 0.0002 |
|  | 2.1073 | 0 |
|  | 2.1417 | 0.0061 |
|  | 2.2226 | 0 |
|  | 2.3314 | 0.0001 |
|  | Σ | **0.0064** |
| IV | 2.0487 | 0.0004 |
|  | 2.0782 | 0.0001 |
|  | 2.0840 | 0.0096 |
|  | 2.1926 | 0 |
|  | 2.3559 | 0.0002 |
|  | Σ | **0.0103** |
| V | 2.0445 | 0.0007 |
|  | 2.0636 | 0.0022 |
|  | 2.0778 | 0.0090 |
|  | 2.1839 | 0.0003 |
|  | 2.3538 | 0.0003 |
|  | Σ | **0.0125** |

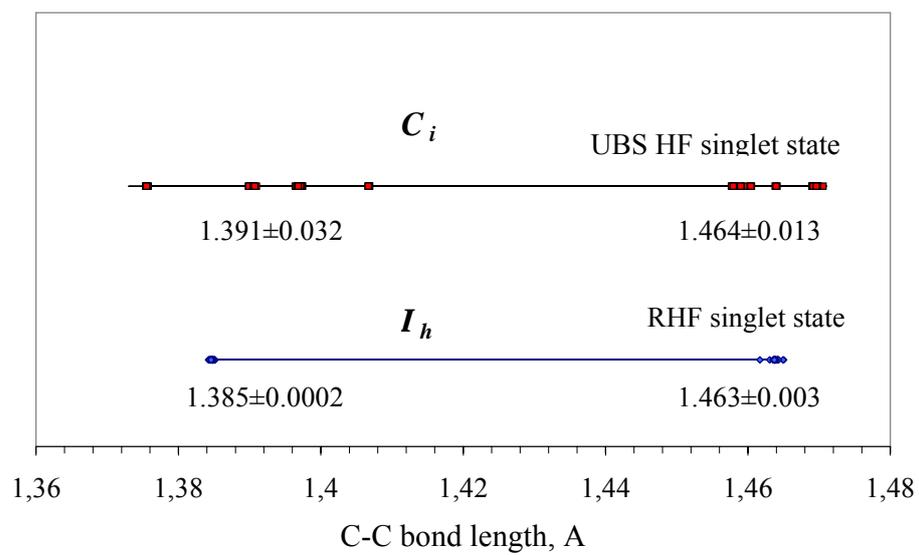

**Figure 1**. Dispersion of the C-C bonds of the $C_{60}$ molecule [15].

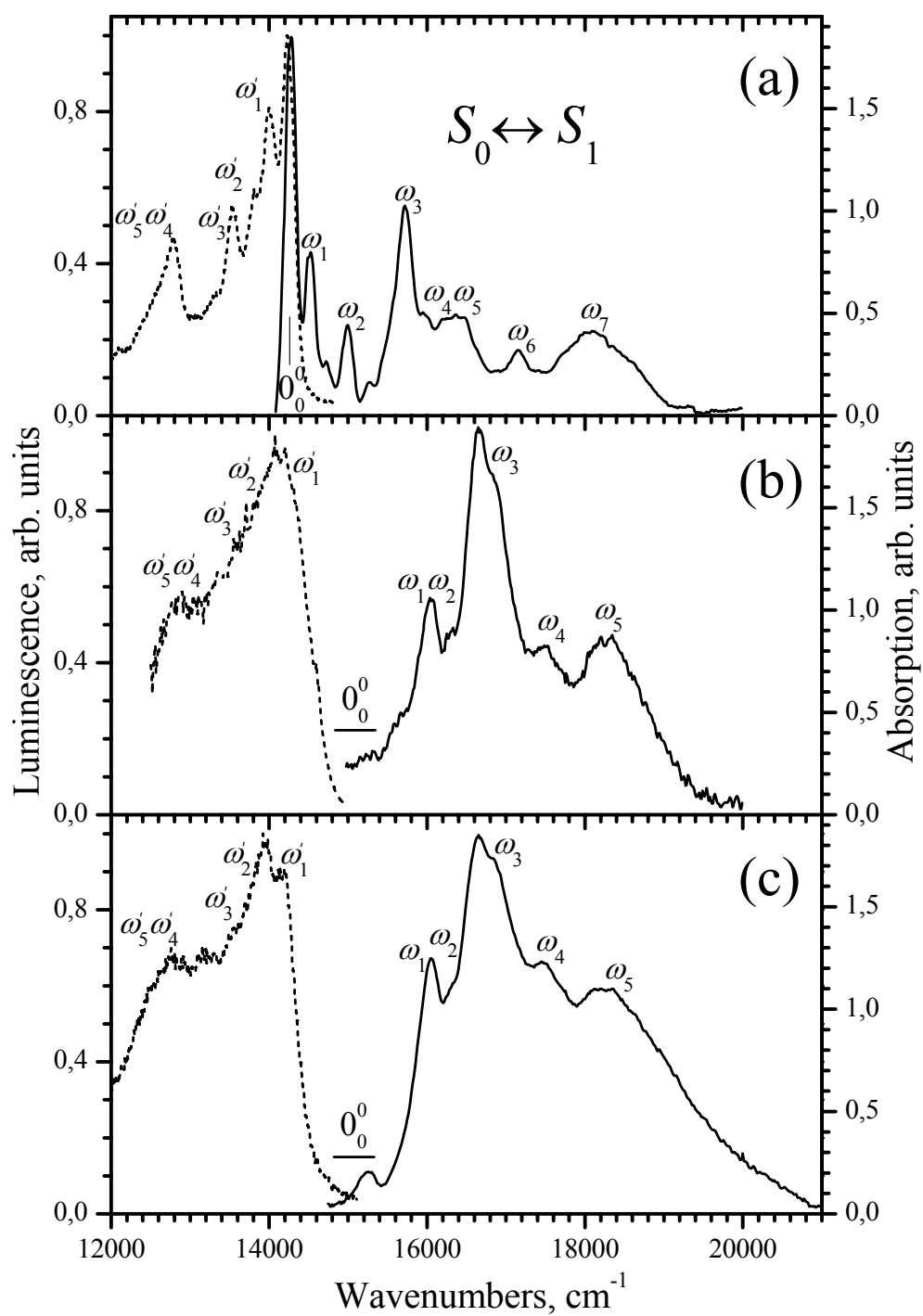

**Figure 2**. Specular background-free absorption (solid lines) and luminescence (dotted lines) spectra of **I** (*a*), **II** (*b*), and $C_{60}$ (*c*) in crystalline toluene (T = 80 K). $\omega_i$ and $\omega'_i$ Mark vibronic bands in the absorption and luminescence spectra, respectively. Vibrational analysis of the vibronic series is presented in [38]. $0^0_0$ Marks the position of a pure electronic transition which is quite certain for **I** and fills some regions for **II** and $C_{60}$ due to multiplet structure.

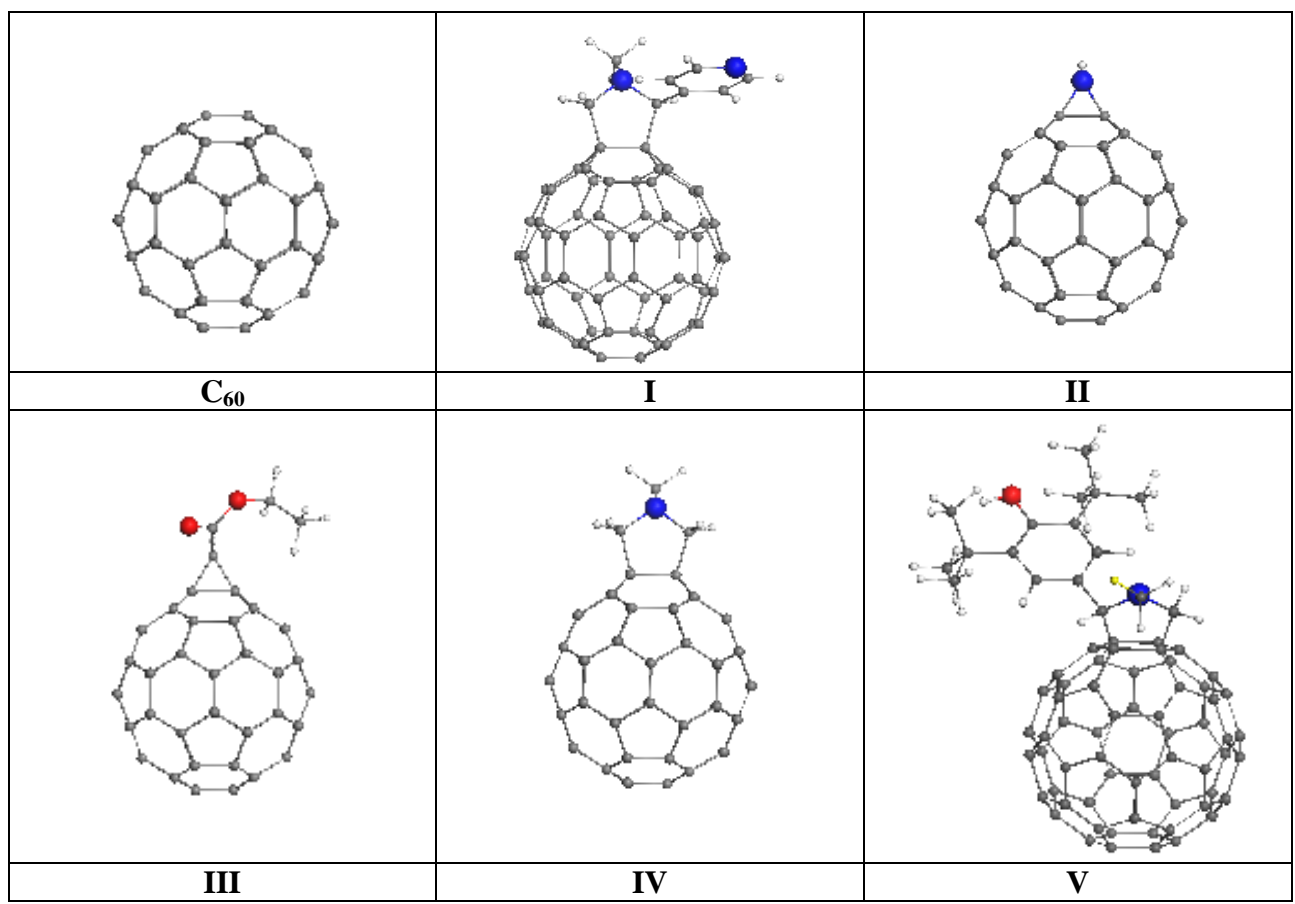

**Figure 3**. Equilibrium structures of fullerene $C_{60}$, N-methyl-2(4-pyridine)-3,4-[$C_{60}$]fulleropyrrolidine (**I**) [38], fullerene azyridine [$C_{60}$] (**II**), ethyl ester of [$C_{60}$]fullerene acetic acid (**III**) [39], N-methyl-3,4-[$C_{60}$]fulleropyrrolidine (**IV**) [39], and N-methyl-2-(3,5-di-*tert*-butyl-4-hydroxyphenyl)-[$C_{60}$] fulleropyrrolidine (**V**) [40]. Carbon atoms are not shown. Big blue and red balls mark nitrogen and oxygen atoms. Hydrogens are shown by small white balls. UBS HF, singlet state.

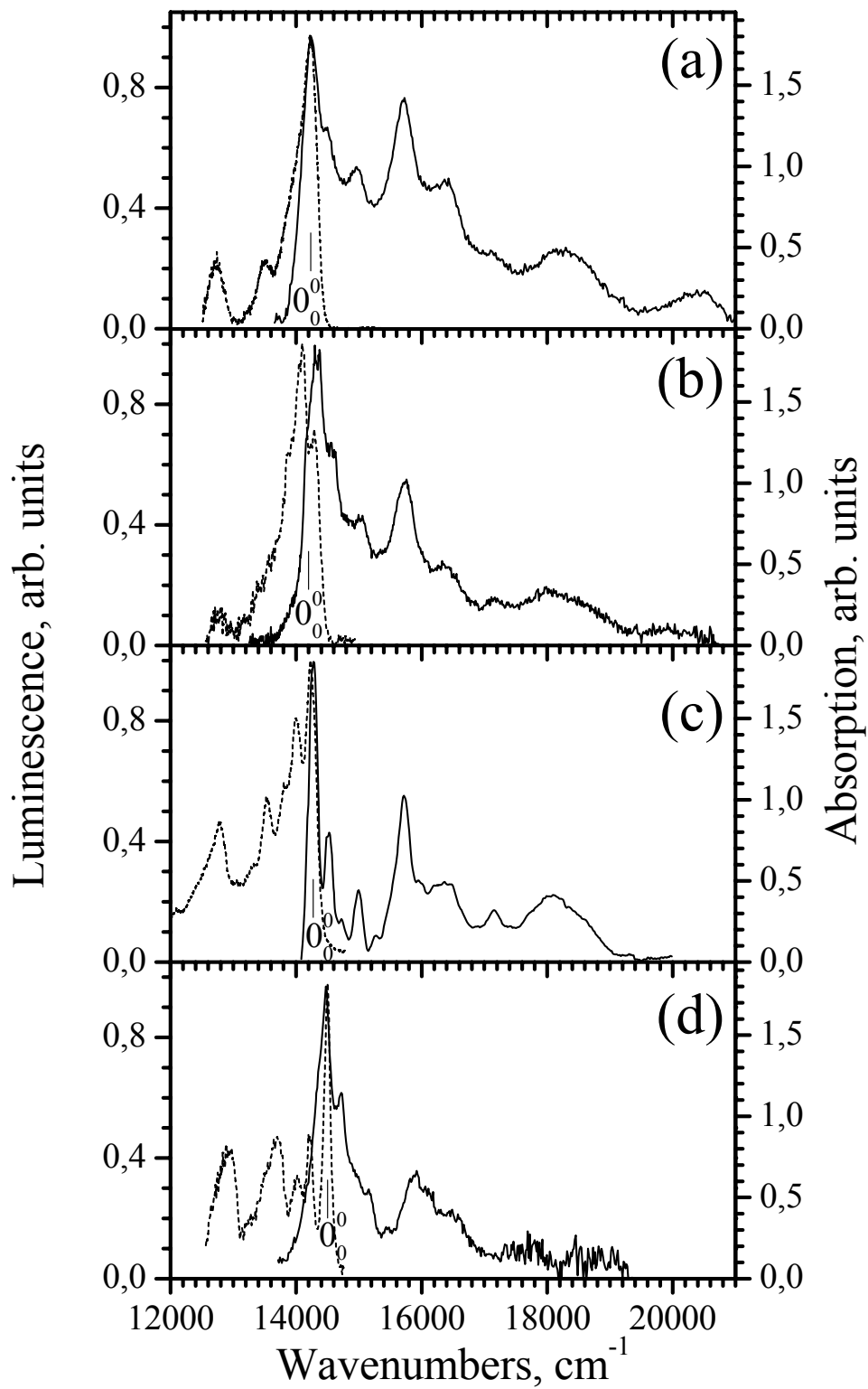

**Figure 4**. Specular background-free absorption (solid lines) and luminescence (dotted lines) spectra of **V** [40] (*a*), **I** (*b*) [38], **IV** [38] (*c*) and **III** [39] (*d*) in crystalline toluene (T = 80 K). $0_0^0$ marks the positions of a pure electronic transitions.

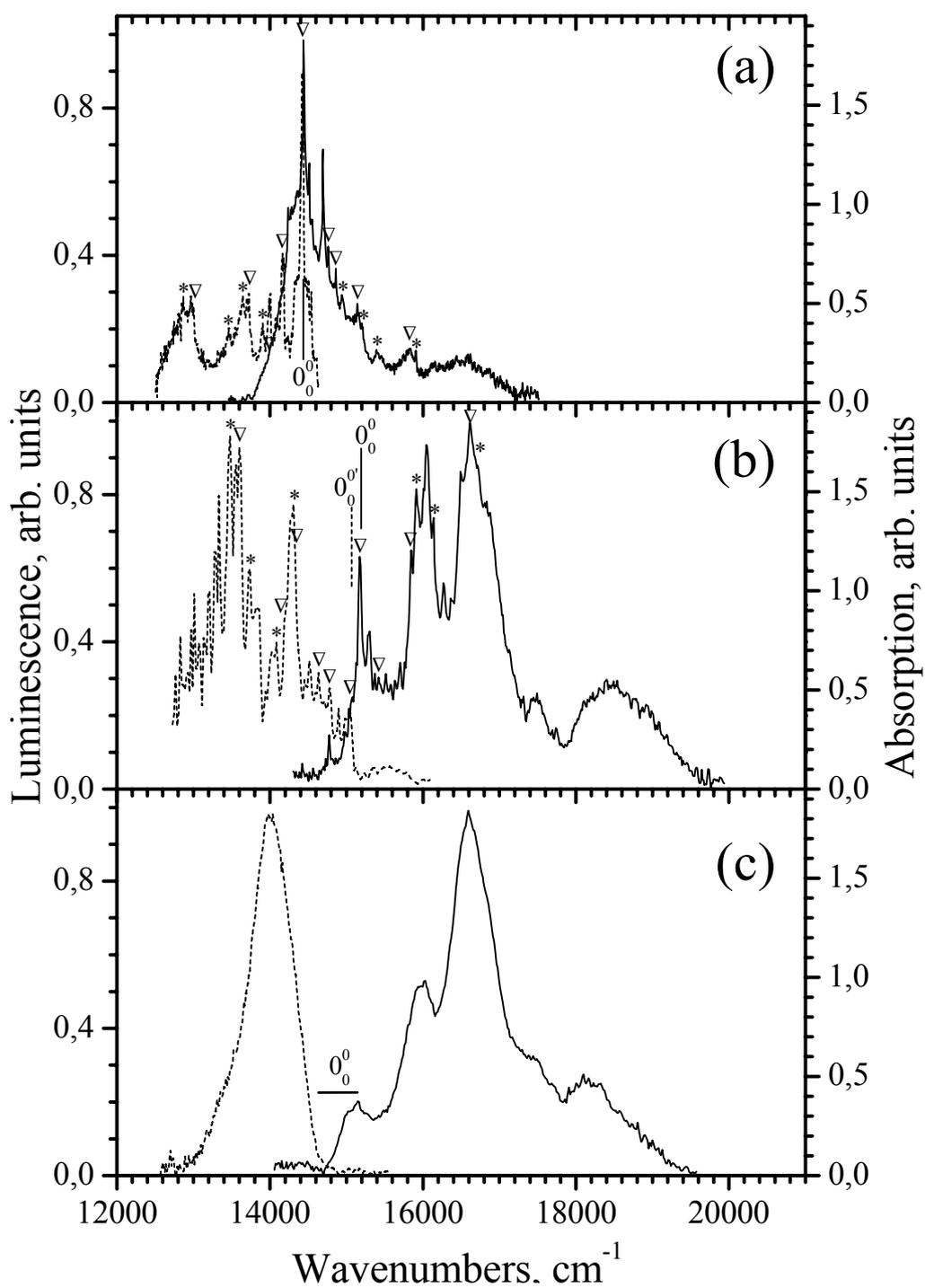

**Figure 5**. Specular background-free absorption (solid lines) and luminescence (dotted lines) spectra of **III** [39] (*a*), $C_{60}$ [38] (*b*), and **II** (*c*) in crystalline toluene (T = 2 K). $0_0^0$ and $0_0^{0'}$ mark the positions of pure electronic transitions. Stars and triangles mark vibronic bands related to the HT and FC series, respectively. The series separation is based on the vibrational analysis of the spectra given in Refs. 38-40.